\documentclass[a4paper]{iopart}
\usepackage{iopams}
\usepackage{graphicx}
\usepackage{psfrag}
\usepackage{times}

\usepackage{mathrsfs}
\DeclareSymbolFontAlphabet{\mathrsfs}{rsfs}
\DeclareMathAlphabet{\mathcal}{OMS}{cmsy}{m}{n}

\newcommand{\scri}{\mathrsfs{I}}

\newcommand{\be}{\begin{equation}}
\newcommand{\ee}{\end{equation}}

\def\tt{{\tilde{t}}}

\def\tg{{\tilde{g}}}

\def\tw{\widetilde{\mathcal{M}}}

\begin{document}


\title{Hyperboloidal evolution with the Einstein equations}

\author{An\i l Zengino\u{g}lu}
\address{Max-Planck-Institut f\"ur Gravitationsphysik,
	     Albert-Einstein-Institut,\\
	     Am M\"uhlenberg~1, D-14476 Golm, Germany}

\begin{abstract}
We consider an approach to the hyperboloidal evolution problem based
on the Einstein equations written for a rescaled metric. It is shown
that a conformal scale factor can be freely prescribed a priori in
terms of coordinates in a well-posed hyperboloidal initial value
problem such that the location of null infinity is independent of the
time coordinate. With an appropriate choice of a single gauge source
function each of the formally singular conformal source terms in the
equations attains a regular limit at null infinity. The suggested
approach could be beneficial in numerical relativity for both wave
extraction and outer boundary treatment.
\end{abstract}


\section{Introduction}

An important problem in general relativity is the calculation of
gravitational radiation emitted by self-gravitating astrophysical
sources. In the isolated system idealization of such objects, one
attaches to the far-field zone an asymptotic region in which the
spacetime becomes flat in a certain sense. The resulting models are
commonly referred to as asymptotically flat spacetimes. Because
gravitational energy is not localizable and there is no generally
satisfactory definition of quasi-local energy available 
\cite{Szabados04}, the concept of gravitational radiation is
rigorously defined in these models only at null infinity
\cite{Bondi62, Sachs62}. As a consequence, one needs global access to
the spacetime solution to discuss gravitational radiation in an
unambiguous way. In numerical calculations, however, one typically
truncates the solution domain by introducing an artificial timelike
outer boundary into the spacetime. This practice introduces certain
well-known conceptual and practical difficulties \cite{Pazos06,
  Rinne07, Buchman07, Sarbach04, Friedrich99, Friedrich:2005kk}.

A clean way to calculate gravitational radiation is to include null
infinity in the computational domain. This can be achieved under
certain conditions by employing the conformal compactification
technique introduced by Penrose which allows the study of null
infinity by local differential geometry \cite{Penrose63, Penrose64,
  Penrose65}.  This framework has been used extensively in the
mathematical discussion of isolated systems \cite{Hawking73, Geroch77,
  Ashtekar80, Bobby82, Friedrich92}. In numerical applications, it has
been successfully applied within the characteristic approach to
spacetimes which can be foliated by null hypersurfaces in a regular
way \cite{Gomez:1994rg, Bartnik99, Gomez:2002ev, Barreto:2004fn,
  Winicour05}. In regions of highly dynamical, strong gravitational
fields, however, characteristic foliations are not well behaved due to
formation of caustics in bundles of light rays generating the null
hypersurfaces \cite{Friedrich83}. While there are promising
suggestions on how to deal with this difficulty such as the
Cauchy-characteristic matching \cite{Bishop93, Bishop96, Bishop96CCM,
  Bishop:1998uk}, a successful implementation has not yet been
achieved.

A more general method to include null infinity in the solution domain
has been suggested by Friedrich \cite{Friedrich83a}. In this approach,
one solves Cauchy problems based on spacelike foliations that
intersect null infinity. Such foliations are called hyperboloidal as
their asymptotic behavior is similar to that of the standard
hyperboloids in Minkowski spacetime. While one solves a Cauchy problem
in the sense of partial differential equations, the underlying
surfaces are not Cauchy surfaces in the sense of differential geometry
because their domain of dependence does not cover the entire
spacetime. To avoid misunderstanding, the related problem is called
the hyperboloidal problem. The hyperboloidal approach is promising as
the underlying surfaces combine favorable properties of standard and
characteristic approaches. They are as flexible as Cauchy-type
foliations commonly used in numerical relativity and they approach
null infinity thus enabling a clean treatment of gravitational
radiation.

In \cite{Friedrich83a}, the Einstein equations are reformulated for a
conformally rescaled metric such that the equations are manifestly
regular at null infinity. The resulting system of regular conformal
field equations is larger than the Einstein equations and involves
evolution equations for the conformal factor. This approach has been
very efficient in the effort to understand the global structure of
spacetimes \cite{Friedrich92, Friedrich86, Friedrich88,
  Friedrich98b}. Numerical work on the hyperboloidal problem for this
system has been performed mainly by H\"ubner \cite{Huebner93,
  Huebner96, Huebner98, Huebner99, Huebner01}, Frauendiener
\cite{Frauendiener98a, Frauendiener98b, Frauendiener98c,
  Frauendiener:2002iw, Frauendiener:2002ix} and Husa
\cite{Husa:2002kk, Husa02b} (see \cite{Frauendiener04} for a
review). Unfortunately, the regular conformal field equations could
not yet be used numerically in the study of highly dynamical systems.

In view of the recent success of numerical codes in solving the
Einstein equations for a large variety of astrophysically interesting
systems \cite{Pretorius05b, Baker05, Campanelli05, Gonzalez:2006md,
  Diener:2005mg, Koppitz:2007ev}, it seems desirable to implement the
hyperboloidal problem directly for the conformal Einstein equations
\cite{Moncrief00, Andersson02a, Husa05, vanMeter:2006mv}. The idea is
the following: Under a conformal rescaling of the metric
$g=\Omega^2\tilde{g}$ with a function $\Omega>0$, the Ricci tensor
transforms as \be\label{eq:ricci} \fl \qquad R_{\mu\nu}[g]=R_{\mu\nu}[\tilde{g}]
- \frac{1}{\Omega}\left(2\, \nabla_\mu \nabla_\nu \Omega + \Box
\Omega\, g_{\mu\nu} \right) + \frac{3}{\Omega^2} (\nabla_\lambda
\Omega)\, \nabla^\lambda \Omega \, g_{\mu\nu}. \ee Here, $\nabla_\mu$
is the Levi-Civita connection of the rescaled metric $g$ and
$\Box:=g^{\mu\nu}\nabla_\mu\nabla_\nu$. The Einstein vacuum field
equations, $G_{\mu\nu}[\tilde{g}]:=R_{\mu\nu}[\tg]
-\frac{1}{2}\tg_{\mu\nu} R[\tg] = 0$, are thus equivalent to a similar
system for the conformally rescaled metric \be \label{eq:comp_einst}
\fl \qquad G_{\mu\nu}[g] = T_{\mu\nu}[\Omega]:= -\frac{2}{\Omega}\left(\nabla_\mu
\nabla_\nu \Omega - \Box\Omega \, g_{\mu\nu}\right) -
\frac{3}{\Omega^2} (\nabla_\lambda \Omega) \nabla^\lambda \Omega\,
g_{\mu\nu}. \ee To include null infinity in the computational domain,
one allows the conformal factor $\Omega$ to vanish in a suitable way
and solves the above system directly. It has been pointed out
\cite{Friedrich02} that there are two major difficulties in this
program, even if one assumes that the conformal extension is
regular. First there is the question of how to fix $\Omega$. The
conformal factor is related to the asymptotic structure of the
spacetime, therefore it must be determined jointly with the metric.
Secondly, there are terms involving divisions by powers of $\Omega$
which are formally singular at $\{\Omega=0\}$. The question then is
how to make sure that these terms attain regular limits at null
infinity.

Assuming a regular conformal extension, these difficulties have been
solved in the characteristic case by Tamburino and Winicour
\cite{Tamburino66}. Their solution includes a certain choice of
coordinate and conformal gauge in which the formally singular terms in
(\ref{eq:comp_einst}) attain regular limits at null infinity. Their
construction provides, in a sense, the mathematical basis for
characteristic codes that include null infinity in the computational
domain. The problem in the hyperboloidal case has been open.

We address the above mentioned difficulties in two propositions. The
first one treats the case of a strictly positive conformal factor and
states that it can be prescribed almost arbitrarily in terms of local
spacetime coordinates in a well-posed Cauchy problem for the Einstein
equations. For the second proposition we allow the conformal factor to
vanish in a certain way so that null infinity can be included in the
computational domain. We see that a class of gauges can be chosen in a
hyperboloidal initial value problem in such a way that the formally
singular terms due to conformal compactification attain regular limits
at null infinity for a certain class of asymptotically flat
spacetimes.

\section{A hyperbolic reduction for a positive conformal factor} \label{wp_red}
The system (\ref{eq:comp_einst}) has the form of Einstein equations
with source terms. In general, such a system must be completed by
additional equations derived from the Bianchi identities, $\nabla^\mu
G_{\mu\nu}=\nabla^\mu T_{\mu\nu}=0$, implying equations of motion for
the source functions. In our case, however, there are no additional
equations required for the conformal factor $\Omega$ as shown below.\\

\textbf{Proposition 1}: \emph{The conformal Einstein equations
  (\ref{eq:comp_einst}) admit a well-posed initial value problem for
  an arbitrary $\Omega\in
  C^3(\mathbb{R}^4,\mathbb{R}_+)$.}

First we show that the Bianchi identities are satisfied for a
positive, sufficiently differentiable conformal factor. We calculate
\begin{eqnarray} \label{contr_tmunu} \nabla^\mu T_{\mu\nu}[\Omega] &=&
-\frac{2}{\Omega^2}\, \nabla^\mu \Omega \, \left(2 \nabla_\mu
\nabla_\nu \Omega + \Box\Omega\, g_{\mu\nu} \right) - \nonumber \\ &-&
\frac{2}{\Omega}\, \left(\Box\nabla_\nu \Omega - \nabla_\nu \Box
\Omega\right) + \frac{6}{\Omega^3}\,(\nabla_\nu \Omega)
(\nabla_\lambda \Omega) \nabla^\lambda \Omega. \end{eqnarray}
Contracting the commutation relation $\nabla_\lambda \nabla_\nu \,
\nabla_\rho \Omega - \nabla_\nu \nabla_\lambda \, \nabla_\rho \Omega =
R_{\lambda\nu\rho}^{\ \quad \sigma} \nabla_\sigma \Omega$ with
$g^{\lambda\rho}$ and exchanging derivatives we get $\Box\nabla_\nu
\Omega - \nabla_\nu \Box \Omega = R_\nu^{\ \sigma}\nabla_\sigma
\Omega$. Using this relation, equation (\ref{eq:comp_einst}) with the
definition of the Einstein tensor $G_{\mu\nu}[g]$ and the conformal
source tensor $T_{\mu\nu}[\Omega]$, we get \be \label{div_tmunu}
\nabla^\mu T_{\mu\nu}[\Omega] = -\frac{2}{\Omega}\,\nabla^\mu \Omega
\, \left(G_{\mu\nu}[g] - T_{\mu\nu}[\Omega]\right) = 0.\ee We see that
the Bianchi identities are satisfied by virtue of the conformal
Einstein equations for a non-vanishing $\Omega$ that is at least three
times differentiable, i.e. $\Omega\in C^3(\mathbb{R}^4,
\mathbb{R}_+)$, but is otherwise arbitrary.  Therefore, $\Omega$ can
be regarded as a free function.  We can write down some suitable
equation for it consistent with the above calculation or prescribe it
directly in terms of some yet unspecified coordinate system as long as
$\Omega\ne 0$.

The next step is to set up a well-posed initial value problem for
(\ref{eq:comp_einst}). We employ the general wave gauge reduction of
the Einstein equations, also known as the generalized harmonic
reduction for historical reasons, where the argument of well-posedness
for (\ref{eq:comp_einst}) can be taken almost directly from
\cite{Friedrich96, Friedrich:2000qv} with only a minor modification.

Regarding the Ricci tensor $R^{\mu\nu}$ as a differential operator
acting on the metric $g$, we can write the conformal Einstein
equations in a local coordinate system $\{x^\mu\}_{\mu=0,1,2,3}$ as
\be \label{eq:comp_ricci} \fl \qquad R^{\mu\nu}[g]=\frac{1}{2}g^{\lambda\rho}
\partial_\lambda\partial_\rho g^{\mu\nu}+ \nabla^{(\mu}\Gamma^{\nu)} -
g^{\lambda\rho}g^{\sigma\tau}\,
\Gamma^\mu_{\lambda\sigma}\Gamma^\nu_{\rho\tau}=T^{\mu\nu} -
\frac{1}{2}g_{\mu\nu} T,\ee where we have defined the contracted
Christoffel symbols $\Gamma^\mu:=g^{\sigma\tau}\Gamma_{\sigma\tau}^\mu
= -\Box\, x^\mu$, set $\nabla^\mu\Gamma^\nu=
g^{\mu\rho}(\partial_\rho\Gamma^\nu+ \Gamma_{\rho\lambda}^\nu
\Gamma^\lambda)$ and $T=g^{\lambda\rho}T_{\lambda\rho}$.  The
principal part of the operator $R^{\mu\nu}$ is of no known type. It
was recognized by Choquet-Bruhat \cite{Choquet52} that one can always
choose a wave gauge, historically referred to as harmonic gauge, at
least locally, so that the contracted Christoffel symbols vanish,
$\Gamma^\mu=-\Box_{g} x^\mu=0$, and the system (\ref{eq:comp_ricci}),
reduces to a quasi-linear system of wave equations. This reduction
technique led to the first local existence result in general
relativity \cite{Choquet52}.

The reduction based on the wave gauge was generalized to arbitrary
coordinate systems by Friedrich with the introduction of gauge source
functions \cite{Friedrich85}. In the general wave gauge, the
coordinates are constructed as solutions to an initial value problem
for the semi-linear system of wave equations $\Box_{g}x^\mu=
-\Gamma^\mu=-F^\mu$ with prescribed functions $F^\mu(x, g)$ that can
depend on the coordinates and the metric. These functions act as
source functions for the coordinate gauge, hence the name gauge source
functions. Note that the general wave gauge, in contrast to the wave
gauge described above, is not a specific choice of coordinates but a
particular way to prescribe general coordinates in an initial value
problem.

The reduced system for (\ref{eq:comp_ricci}) is then obtained by
replacing the contracted Christoffel symbols with the gauge source
functions $F^\mu$.  The result is a quasi-linear system of wave
equations for the metric components which can be written as
\be\label{reduced_einst} G^{\mu\nu}[g] = T^{\mu\nu}[\Omega] +
\nabla^{(\mu} C^{\nu)} - \frac{1}{2} (\nabla_\lambda C^\lambda)
g^{\mu\nu}, \ee where $C^\mu=\Gamma^\mu-F^\mu$ are called the
constraint fields. We want to study the Cauchy problem for this
system.  We will only point out certain aspects that play a role in
later considerations or that are different from the detailed
discussion in \cite{Friedrich:2000qv}.

The Cauchy data on an initial hypersurface
$\mathcal{S}\equiv\{x^0=0\}$ consist of $g^{\mu\nu}|_{\mathcal{S}}$
and $\partial_0 g^{\mu\nu}|_{\mathcal{S}}$.  Assume we are given on
$\mathcal{S}$ a Riemannian metric $h_{ab}$ and a symmetric tensor
field $K_{ab}$ as a solution to the Einstein constraint equations
where $a,b=1,2,3$. We choose gauge source functions $F^\mu(x^\lambda)$
and four functions on $\mathcal{S}$ that correspond to initial data
for the lapse function $\alpha>0$ and the three components of the
shift vector $\beta^a$. In the interior, these functions should be
chosen such that $\partial_0$ is timelike which implies
$\alpha^2-h_{ab}\beta^a \beta^b>0$. We will later allow $\partial_0$
to become null at the outer boundary (see the discussion leading to
(\ref{eq:lapse_shift})). We obtain the data
$g^{\mu\nu}|_{\mathcal{S}}$ via the decomposition \be\label{eq:deco} g
= g^{\mu\nu} \partial_\mu \partial_\nu =
-\frac{1}{\alpha^2}\partial_0^2 + \frac{2}{\alpha^2}\beta^a \partial_0
\partial_a + \left(h^{ab}- \frac{\beta^a
  \beta^b}{\alpha^2}\right)\partial_a\partial_b.\ee The data
$\partial_0g^{\mu\nu}|_{\mathcal{S}}$ is determined such that
$C^\mu|_{\mathcal{S}}=0$ and $K_{\alpha\beta}$ is the second
fundamental form on $\mathcal{S}$.  Standard theorems guarantee that
we can find a unique solution to the Cauchy problem for the reduced
equations (\ref{reduced_einst}) that depends continuously on the
initial data.  The solution spaces of (\ref{reduced_einst}) and
(\ref{eq:comp_einst}) are equivalent if the constraint fields
vanish. The Bianchi identity, $\nabla_\mu G^{\mu\nu}=0$, and
(\ref{div_tmunu}) together with (\ref{reduced_einst}) imply the
following subsidiary system for the constraint fields
\be \label{sub_sys} \Box C^\mu + R^{\mu}_{\nu}C^\nu
-\frac{4}{\Omega}\nabla_\nu \Omega \left(\nabla^{(\mu} C^{\nu)} -
\frac{1}{2} (\nabla_\lambda C^\lambda) g^{\mu\nu}\right)=0.\ee Initial
data for the evolution equations has been constructed such that
$C^\mu|_{\mathcal{S}}=0$. From the evolution equations evaluated on
$\mathcal{S}$ it follows that $\partial_0 C^\mu|_{\mathcal{S}}=0$.
The uniqueness of solutions to the Cauchy problem for the semi-linear,
homogeneous system of wave equations for $C^\mu$ given in
(\ref{sub_sys}) then implies that the solution to the reduced system
(\ref{reduced_einst}) satisfies $C^\mu=0$ away from the initial
surface $\mathcal{S}$. Thus we have shown that there is a well-posed
hyperbolic reduction for the conformal Einstein equations
(\ref{eq:comp_einst}).\\

We shall briefly elaborate on how the free prescription of the
conformal factor is to be understood. We can not prescribe the
conformal factor as a function on the manifold because, in an initial
value problem, we do not know the manifold. The prescription of a
function for the conformal factor determines only its representation
in terms of coordinates which are yet to be constructed during the
solution process. Invariant properties of the resulting conformal
factor will depend on initial data and the choice of gauge source
functions.  The essential property of (\ref{eq:comp_einst}) that is
responsible for this feature is its conformal invariance, in the sense
that if $(\tw,g,\Omega)$ is a solution to $G_{\mu\nu}[g] =
T_{\mu\nu}[\Omega]$, then $(\tw,\omega^2 g,\omega \Omega)$ with a
positive function $\omega$ is a solution to $G_{\mu\nu}[\omega^2 g] =
T_{\mu\nu}[\omega\Omega]$.  The system (\ref{eq:comp_einst})
determines the conformal class of the metric $g$ in contrast to the
Einstein equations which determine the isometry class of
$\tilde{g}$. This allows us to prescribe an arbitrary coordinate
representation for the conformal factor as long as there are no
geometric requirements to be satisfied.


\section{Choice of gauge at null infinity}\label{sec:gauge}
In this section, we deal with the formally singular terms in the
conformal source tensor $T_{\mu\nu}[\Omega]$. We first present a
preferred conformal gauge at null infinity in which each of the
conformal source terms attains a regular limit at null infinity in a
given conformal extension. Then we present how this choice of gauge
can be achieved by a suitable choice of gauge source functions in a
hyperboloidal initial value problem where the conformal factor has
been prescribed explicitly in terms of coordinates.  We restrict our
attention to future null infinity, denoted by $\scri^+$. The treatment
of past null infinity follows by time reversal.

\subsection{The preferred conformal gauge at $\scri^+$}\label{sec:pref_gauge}
Assume that a solution $(\tw,\tg)$ to the Einstein vacuum field
equations has been given which admits a conformal extension
$(\mathcal{M},g,\Omega)$ including a smooth piece of $\scri^+$. The
existence of a broad class of such solutions is due to
\cite{Friedrich83a, ACF92}. It has also been shown that solutions
exist which not only admit a smooth piece of $\scri^+$ but a complete
$\scri^+$ \cite{Chrusciel-Delay-2002} (see also \cite{Friedrich86,
  Corvino-Schoen, Friedrich04}). We show that $\scri^+\subset
\{\Omega=0,d\Omega\ne0\}$ is a shear-free null surface independently
of the conformal gauge \cite{Penrose65, Stewart}. Multiplying
(\ref{eq:comp_einst}) with $\Omega^2$ and evaluating it along
$\scri^+$ we see that
$g^{\lambda\rho}\nabla_\lambda\Omega\nabla_\rho\Omega|_{\scri^+}=0$. This
shows together with $d\Omega|_{\scri^+}\ne 0$ that $\scri^+$ is a null
surface. Now multiply (\ref{eq:comp_einst}) with $\Omega$ and take its
trace-free part along $\scri^+$ to get \be \label{shear_free}
\left(\nabla_\mu \nabla_\nu \Omega-\frac{1}{4} g_{\mu\nu}\, \Box
\Omega\right)\Big|_{\scri^+} = 0, \ee The relation above is
independent of the conformal gauge because we derived it from the
conformal transformation behavior of the Einstein tensor
(\ref{eq:comp_einst}). Another way to see the conformal invariance of
(\ref{shear_free}) is to consider the transformation behavior of
(\ref{shear_free}) under a further rescaling of the conformal metric
given by \be\label{con_res} g'=\omega^2 g,\quad \Omega'=\omega
\Omega,\quad \omega>0 \ \textrm{on}\ \mathcal{M}.\ee We have
$\nabla'_\mu \nabla'_\nu\Omega'|_{\scri^+} = \omega \nabla_\mu
\nabla_\nu \Omega+ g_{\mu\nu}\, \nabla^\lambda\Omega
\nabla_\lambda\omega$. The trace of this relation reads
\be\label{box_trafo}\Box' \Omega'|_{\scri^+}= \frac{1}{\omega^2}
\left(\omega\Box\Omega+ 4\, \nabla^\lambda\Omega \nabla_\lambda\omega
\right), \ee whence we get
\[\left(\nabla'_\mu \nabla'_\nu \Omega' - \frac{1}{4}g'_{\mu\nu}\,
\Box' \Omega'\right)\Big|_{\scri^+} =\omega\,\left(\nabla_\mu \nabla_\nu \Omega -
\frac{1}{4}g_{\mu\nu}\,\Box \Omega\right)\Big|_{\scri^+}=0. \]

To see that (\ref{shear_free}) implies shear-freeness of $\scri^+$, we
introduce in a neighborhood of $\scri^+$ a null vector field $l^\mu$
that satisfies $l^\mu|_{\scri^+} = \nabla^\mu \Omega$. We complete
$l^\mu$ to a Newman-Penrose complex null tetrad $(l,k,m,\bar{m})$
satisfying the usual relations \cite{Newman62a}.  Newman and Penrose
introduce twelve complex functions called spin coefficients. We are
interested in two of them, namely $\sigma := m^\mu m^\nu \nabla_\mu
l_\nu$ and $\rho := m^\mu \bar{m}^\nu \nabla_\mu l_\nu$. As discussed
in \cite{Newman62a}, when $l^\mu$ is tangent to an affinely
parametrized null geodesic, $\sigma$ can be interpreted as the complex
shear of the null geodesic congruence given by $l^\mu$ and the
expansion of the congruence is characterized by $\rho$.  We see with
(\ref{shear_free})
\[\sigma|_{\scri^+} = \frac{1}{4}\,m^\mu m^\nu g_{\mu\nu} \Box \Omega
= 0,\quad \rho|_{\scri^+} =\frac{1}{4}\,m^\mu \bar{m}^\nu g_{\mu\nu}
\Box \Omega = \frac{1}{4}\Box \Omega. \] In our case, the null
generators of $\scri^+$ are not necessarily geodesic, that is, in
general they do not satisfy $l^\lambda\nabla_\lambda l^\mu=0$ on
$\scri^+$. However, under a rescaling of $l^\mu$ given by $(l')^\mu=
\theta l^\mu$ with a positive function $\theta$, the spin coefficient
$\sigma$ transforms as $\sigma'=\theta \sigma$, so the vanishing of
$\sigma$ is invariant under such a rescaling which we can use to make
$l^\mu$ geodesic. We conclude that $\scri^+$ is a shear-free surface
in any conformal gauge.

While the vanishing of $\sigma$ and thus the shear-freeness of
$\scri^+$ is conformally invariant, the vanishing of the expansion of
$\scri^+$, characterized by $\Box\Omega|_{\scri^+}$, depends on the
conformal gauge. As can be seen from (\ref{box_trafo}), given a
conformal extension we can always find a rescaling (\ref{con_res})
such that $\Box'\Omega'|_{\scri^+}=0$ by solving the ordinary
differential equation \mbox{$\nabla^\lambda\Omega\, \nabla_\lambda
  \ln\omega|_{\scri^+}= -\frac{1}{4}\Box\Omega|_{\scri^+}$}.

We call the conformal gauge in which the expansion of $\scri^+$
vanishes a \textit{preferred conformal gauge}. This gauge has been
useful in mathematical studies because of its special properties
\cite{Stewart, Penrose84}. It is also the gauge choice of Tamburino and
Winicour \cite{Tamburino66}.  By (\ref{shear_free}) and
(\ref{eq:comp_einst}) we see that in a preferred conformal gauge
\[\nabla_\mu \nabla_\nu \Omega |_{\scri^+}=0, \quad \mathrm{and}\quad
\lim_{\Omega\to 0}\,\frac{1}{\Omega}\,g^{\lambda\beta}\nabla_\lambda\Omega
\nabla_\beta\Omega=0,\] which implies that each conformal source
term in (\ref{eq:comp_einst}) attains a regular limit at
$\scri^+$.
\subsection{The choice of gauge source functions at $\scri^+$}
The above construction of a preferred conformal gauge assumes that a
conformal extension has been given. We are interested, however, in the
case where only initial data has been given and we would like to know
how to choose the gauge source functions suitably so that the
hyperboloidal evolution is performed in a preferred conformal gauge.\\

\textbf{Proposition 2}: \emph{Assume hyperboloidal initial data has
  been given whose evolution admits a smooth conformal
  compactification at $\scri^+$. A preferred conformal gauge can be
  achieved in a general wave gauge reduction of (\ref{eq:comp_einst})
  by using $\Omega$ as a coordinate near $\{\Omega=0\}$ and choosing
  the related gauge source function $F^{\Omega}$ such that
  $F^{\Omega}|_{\{\Omega=0\}}=0$.}

We can use the conformal factor as a coordinate near $\{\Omega=0\}$
because \mbox{$d\Omega|_{\{\Omega=0\}}\ne 0$}. In the general wave gauge,
there is a gauge source function related to each coordinate via
$\Box_g x^\mu = -F^\mu$. Using the conformal factor as a coordinate,
we see that the value of the gauge source function
$F^\Omega=-\Box\Omega$ at $\scri^+$ is a direct measure of the
expansion of $\scri^+$. Setting it to zero makes the expansion vanish
and hence gives a preferred conformal gauge.\\

While its proof is very simple, proposition 2 is a remarkable property
of the conformal Einstein equations. In general, it is not known how a
desirable gauge can be achieved by a suitable choice of gauge source
functions \cite{Lindblom:2007xw} because geometric properties of
coordinates depend not only on this choice but also on initial data in
an essentially non-linear way. The above proposition is very special
in this respect as it states that one can, by a suitable choice of a
single gauge source function, fix an a priori known coordinate surface
to be a null surface free of shear and expansion that corresponds to
null infinity.

In a practical numerical calculation one may proceed as follows.  One
calculates on a three dimensional surface $\mathcal{S}$ with a two
dimensional boundary $\Sigma$ hyperboloidal initial data
$(\mathcal{S}, h_{ab}, K_{ab})$ whose evolution admits a smooth
conformal compactification at $\scri^+$. The existence of such data
has been studied in \cite{ACF92}. In the case of isotropic extrinsic
curvature, that is, under the assumption $K_{ab}=\frac{K}{3}\,
h_{ab}$, the data needs to be such that the trace-free part of the
second fundamental form induced by $h_{ab}$ on $\Sigma$ vanishes. This
condition corresponds to shear-freeness of $\scri^+$. The case with
more general data has been considered in \cite{Andersson-Chrusciel93,
  AMF94}. One can construct the data in such a way that the conformal
factor has the form $\Omega=1-r$ in a neighborhood of $\Sigma$ where
$r$ is a coordinate whose level surfaces have spherical topology
\cite{Andersson02a}. The form of the conformal factor in the interior
can be chosen freely to satisfy practical needs, for example to obtain
high resolution in certain domains. Initial data for lapse and shift
should be chosen such that $ \Omega^\mu|_{\scri^+}= -g^{\mu r}|_{r=1}
= \partial_0^\mu = \delta^\mu_0$, which implies with (\ref{eq:deco})
\be\label{eq:lapse_shift} \alpha|_{\Sigma} =\sqrt{h^{rr}},\qquad
\beta^{a}|_{\Sigma}=-h^{a r}.\ee The gauge source functions can be
given quite freely as long as the condition $F^r|_{r=1}=0$ is
satisfied. Then each of the conformal source terms in
\[ T_{\mu\nu} = - \frac{2}{1-r}(\Gamma_{\mu\nu}^r
- g_{\mu\nu}F^r) - \frac{3}{(1-r)^{2}}g_{\mu\nu}g^{rr}, \] will attain
a regular limit at $\scri^+$ and the system (\ref{eq:comp_einst}) can,
in principle, be solved. In practice, however, the accurate
calculation of these limits poses a major computational difficulty
that needs further study.

\section{Discussion}
We studied the hyperboloidal evolution problem for the conformal
Einstein equations. Two propositions have been made that suggest an
alternative method for numerical calculations of asymptotically simple
spacetimes including null infinity, substantiating earlier
expectations in this direction \cite{Moncrief00, Andersson02a, Husa05,
  vanMeter:2006mv}.

Proposition 1 states that one can formulate a well-posed initial value
problem for the Einstein equations in which a positive conformal
factor is prescribed freely in terms of coordinates. In a sense, this
result can be regarded as trivial because a rescaling with a positive
factor essentially amounts to a relabeling of the metric. On the other
hand, the possibility of prescribing the conformal factor in a
well-posed initial value problem is very convenient because it allows
us to set up a formulation of the Einstein equations for a rescaled
metric in which the conformal factor is independent of the time
coordinate in the whole solution domain. Depending on practical
demands, one can choose the conformal factor in certain regions to be
unity so that the standard Einstein equations are obtained, or larger
than unity for refinement.

Proposition 2 shows how one can construct the solution to a
hyperboloidal initial value problem directly in a preferred conformal
gauge by a suitable choice of a single gauge source function. In this
gauge the conformal source terms attain regular limits at null
infinity. In combination with Proposition 1, this allows us to include
null infinity in the solution domain in such a way that the outer
boundary coincides with null infinity. It is a remarkable feature of
the conformal Einstein equations in the general wave gauge that the
constraint equations, the evolution system, the gauge system and the
subsidiary system all work in harmony with conformal transformation
formulae allowing us to make the geometric properties of null infinity
manifest on an a priori known coordinate surface.

We have assumed vacuum throughout the calculations. Matter can be
included by solving the matter equations with respect to the
conformally rescaled metric if the matter fields remain confined to a
finite region of space or fall off sufficiently fast towards infinity.

The approach presented in this article has various advantages that are
favorable for numerical calculations. First of all, the method is
tailored for the treatment of the asymptotic region such that the
interior scheme is not necessarily modified. In fact, hyperboloidal
foliations can be made to coincide with Cauchy-type foliations in the
interior where the conformal factor can be set to unity if desired
\cite{Zeng07b}.  In addition, the principal part of the equations to
be solved in the exterior domain is identical with the standard
Einstein equations which we know how to treat numerically for a large
class of astrophysically interesting configurations. Furthermore, the
location of null infinity is known a priori simplifying numerical
outer boundary treatment and wave extraction significantly. The
calculation of the news function is in our coordinates very simple,
because the solution is obtained directly in Bondi coordinates at null
infinity in contrast to the characteristic approach where the freedom
in the null coordinates is fixed in the interior of the spacetime
\cite{Bishop:1997ik, Bishop:2003bz}.

A successful implementation of the hyperboloidal approach would solve
two major problems in numerical relativity, namely the outer boundary
problem and the wave extraction problem.  One should be aware,
however, that many promising ideas appeared during the history of
numerical relativity which encountered major difficulties in practical
implementation that remain unsolved. In our case the most delicate
issue is whether the formally singular terms can be numerically
calculated in a stable manner. This question is open and is left for
future work. We emphasize, however, that the viability of a similar
calculation has been demonstrated within the characteristic approach
long time ago \cite{Isaacson:1983xc}.

It should be pointed out that there is a theoretical limitation to the
hyperboloidal approach. It does not allow us to calculate global
spacetimes because spatial infinity is not included in the
computational domain. A detailed study of the asymptotic behavior of
gravitational fields near spatial infinity is difficult and an active
area of research \cite{Friedrich04, Friedrich98, ValienteKroon04,
  ValienteKroon04a, ValienteKroon06}. It seems that one should first
achieve a successful implementation of the hyperboloidal problem for
astrophysically interesting configurations before attacking the more
challenging problem of spatial infinity.

\section*{Acknowledgments}
I thank Sascha Husa for suggesting the research project. I am grateful
to Helmut Friedrich and Sascha Husa for guidance into the conformal
approach and comments on the manuscript. I would also like to thank
Robert Beig, Carsten Schneemann and Jeffrey Winicour for discussions.
\section*{References}

\bibliographystyle{iopart-num}
\bibliography{references}

\end{document}